\documentclass[12pt]{article}

\catcode`\@=11

\global\arraycolsep=2pt
\usepackage{amsbsy,amssymb,latexsym,amsfonts,amsmath}
\usepackage{graphicx,color}

\allowdisplaybreaks

\begin{document}

\vspace*{0.7cm}

\begin{center}
{ \Large Spontaneously broken supersymmetric fracton phases with fermionic subsystem symmetries}
\vspace*{1.5cm}\\

Hosho Katsura${}^{1,2,3}$ and Yu Nakayama${}^4$ \\
\vspace{.2 in}
${}^{1}$\textit{Department of Physics, The University of Tokyo, Tokyo, Japan}

\vskip 1pt
\vspace{.1 in}
${}^{2}$\textit{Institute for Physics of Intelligence, The University of Tokyo, Tokyo, Japan}

\vskip 1pt
\vspace{.1 in}
${}^{3}$\textit{Trans-scale Quantum Science Institute, University of Tokyo, Tokyo, Japan}

\vskip 1pt
\vspace{.1 in}
${}^{4}$\textit{Department of Physics, Rikkyo University, Tokyo, Japan}


%


%

%

\vspace{3.8cm}
\end{center}

\begin{abstract}
We construct a purely fermionic system with spontaneously broken supersymmetry that shares the common feature with a fracton phase of matter. Our model is gapless due to the Nambu-Goldstone mechanism. It shows a ground-state degeneracy with the ``Area-law" entropy due to fermionic subsystem symmetries.  In the strongly coupled limit, it becomes a variant of the Nicolai model, and the ground-state degeneracy shows the ``Volume-law" entropy.
Gauging the fermionic subsystem symmetry has an t'Hooft anomaly by itself, but the would-be gauged theory may possess a fermionic defect that is immobile in certain spatial directions.
\end{abstract}

\thispagestyle{empty} 

\setcounter{page}{0}

\newpage

\section{Introduction}
Ordinary matter has extensive entropy at finite temperature, which is proportional to the whole volume of the system, while it has zero entropy at zero temperature, satisfying the third law of the thermodynamics. Certain peculiar matter (e.g. in the glassy phase) shows non-zero entropy at zero temperature that is proportional to the volume.
There exists, however, even more peculiar matter which shows the ``Area-law" entropy at zero temperature. One of the most well-known, yet mysterious, such examples is a black hole \cite{Bekenstein:1973ur}.\footnote{Uncharged black holes have zero entropy at zero temperature satisfying the third law of thermodynamics, but charged black holes have zero-temperature entropy that is proportional to the area of the horizon.} Another example is a ground-state entropy of the fracton phase of matter 
\cite{Chamon:2004lew, Vijay:2015mka, Vijay:2016phm, Pretko:2016kxt}. 
A fracton is an extraordinarily peculiar excitation and has attracted a lot of attention these days. See e.g. \cite{Nandkishore:2018sel,Pretko:2020cko,Grosvenor:2021hkn} for reviews.

In the fracton phase of matter, the origin of the ``Area-law" entropy can be attributed to the so-called subsystem symmetry \cite{sub00,sub01,sub02,sub03,sub1,sub2,sub3,sub4,sub5,sub6,Rayhaun:2021ocs}.
The Noether charge for the ordinary symmetry (in $d$ dimensional space) is obtained by integrating the charge density over the whole volume: $Q_0= \int d^d x\, \rho$ while that for the subsystem symmetry is obtained by integrating the charge density only over codimension $q$ subsystems $Q_q = \int d^{d-q} x\, \rho$. A subsystem symmetry can be regarded as infinitely many copies of ordinary symmetries and it results in a huge degeneracy in the energy spectrum of the system.

Although many realizations of subsystem symmetries have been studied (both in lattice models and in continuum field theories) in the context of fracton physics, a construction of  a fermionic fracton has been a challenging issue. In  \cite{Yamaguchi:2021qrx}, they proposed a supersymmetric field theory that has a fermionic subsystem symmetry. In this paper, we construct a purely fermionic system with spontaneously broken supersymmetry that shares the common feature with a fracton phase of matter\footnote{Note that our effective field theory description, similar to those in \cite{PBF,Seiberg:2020bhn,Seiberg:2020wsg,Yamaguchi:2021qrx,Burnell:2021reh,Gorantla:2021bda}, will not include a ``fracton" as a dynamical degree of freedom, but it appears as a probe defect.}.

Our model has one parameter, and the weak coupling limit gives a tight-binding model of free fermions with a peculiar dispersion relation $E = |k_x k_y|$ with respect to the wavenumber $k_x$ and $k_y$. In the strongly coupled limit, our model becomes a two-dimensional variant of the Nicolai model \cite{Nicolai:1976xp}. The Nicolai model is a supersymmetric lattice model whose Hamiltonian is constructed purely out of fermionic creation and annihilation operators. The ground states on the chain are exponentially degenerate with respect to the system size, and the structure has been extensively studied in the literature \cite{Katsura:2017pjd,Moriya:2016zqu,La:2018zkj}. When the coupling constants are randomly generated, it may also be regarded as a supersymmetric analogue of the Sachdev-Ye-Kitaev model \cite{Sannomiya:2016mnj,Gross:2016kjj,Fu:2016vas,Iyoda:2018osm}. Our model therefore interpolates a fracton phase of matter and a higher-dimensional variant of the Nicolai model.

The organization of the paper is as follows. In section 2, we construct a purely fermionic system with spontaneously broken supersymmetry that shares the common feature with a fracton phase of matter as a continuum field theory. In section 3, we present a lattice regularization and study the ground state degeneracy. In section 4, we discuss some related models. In section 5, we study gauging of the fermionic subsystem symmetry in relation to the defect operators. In section 6, we conclude with some discussions. The paper has two appendices. In Appendix A, we review the superfield formalism, and in Appendix B, we present some general aspects of gauging shift symmetries.

\section{Continuum Model}
Let us consider a continuum field theory in $d=1+2$ dimensions $(t,\vec{x}) = (t,x,y)$ with a complex supercharge
\begin{align}
Q  = \int d^2x\, \left( g \chi + \chi \partial_x \chi \partial_y \psi \right) , \label{Q}
\end{align}
where $\chi$ and $\psi$ are complex fermions with the canonical (equal-time) anti-commutation relations: $\{\chi^\dagger(\vec{x}), \chi(\vec{x}')\} = \{\psi^\dagger(\vec{x}), \psi(\vec{x}')\} = \delta(\vec{x}-\vec{x}')$ and $\{\chi(\vec{x}), \chi(\vec{x}')\} = \{\chi(\vec{x}), \psi(\vec{x}')\} = \{\psi(\vec{x}), \psi(\vec{x}')\} = \{\chi^\dagger(\vec{x}), \psi(\vec{x}') \}= 0  $.

We define the supersymmetric Hamiltonian as the anti-commutator of the complex supercharges:
\begin{align}
H &= Q^\dagger Q + Q Q^\dagger \cr
& = \int  d^2x\, \left( g^2 + g\left( \chi^\dagger \partial_x \partial_y \psi^\dagger - \chi \partial_x \partial_y \psi \right) + H_{\mathrm{int}}  \right) \ , 
\end{align}
where $H_{\mathrm{int}}$ the four-fermi interaction: 
\begin{align}
H_{\mathrm{int}} = -\int d^2x\, & \Big( \partial_y (\chi^\dagger \partial_x \chi^\dagger) \partial_y (\chi \partial_x \chi) + (\partial_x \chi \partial_y \psi)(\partial_x \chi^\dagger \partial_y \psi^\dagger) \cr 
+& \partial_x(\chi \partial_y \psi)(\partial_x \chi^\dagger \partial_y \psi^\dagger) + \partial_x (\chi^\dagger \partial_y \psi^\dagger)(\partial_x \chi \partial_y \psi) + \partial_x(\chi \partial_y \psi) \partial_x (\chi^\dagger \partial_y \psi^\dagger) \Big)
\end{align}
so that the supercharges are conserved $[H,Q] = [H,Q^\dagger] = 0 $. The conservation is a direct consequence of the nilpotent property of the supercharges: $\{Q, Q\} = \{Q^\dagger, Q^\dagger\} = 0$, which can be shown from \eqref{Q}. 

By construction, this model is supersymmetric although we have only fermionic canonical fields. It turns out that the supersymmetry $Q$ (as well as $Q^\dagger$) is spontaneously broken when $g\neq0$. There, a superpartner of a fermionic particle is a pair of the fermionic particle itself and a Numbu-Goldstnoe fermion.

This model has a fermionic subsystem symmetry (or kinematic supersymmetry):
\begin{align}
\psi(\vec{x}) \to \psi(\vec{x}) + \xi (x) \ , \label{subsystem}
\end{align}
where the fermionic parameter $\xi(x)$ depends only on the $x$ coordinate but not on the $y$ coordinate.
The corresponding charges are given by
\begin{align}
q^\dagger_y(x) &= \int dy\, \psi^\dagger (\vec{x})  \ , \label{number}
\end{align}
where the integration is over only the $y$ coordinate. These charges  are parametrized by the coordinate $x$, and they are all conserved because $\partial_t{\psi} = \partial_y (...)$ from the equations of motion. In the infinite $g$ limit, our model has another fermionic subsystem symmetry:
\begin{align}
\psi(\vec{x}) \to \psi(\vec{x}) + \zeta(y) \ , \label{dipole}
\end{align}
where the fermionic parameter $\zeta(y)$ depends on the $y$ coordinate but not on the $x$ coordinate. This subsystem symmetry, however, is explicitly broken by the four-fermi interaction $H_{\mathrm{int}}$.

This model also possesses the $U(1)$ fermion number symmetry:
\begin{align}
\chi &\to e^{i\phi} \chi \ ,\cr
\psi &\to e^{-i\phi} \psi \ .
\end{align}
It is generated by the Noether charge
\begin{align}
Q_{U(1)} = \int d^2x \left( \chi^\dagger \chi - \psi^\dagger \psi\right) \ . \end{align}

Other discrete symmetries include $x$-parity symmetry: $x \to - x$, $y \to y$ with $\chi \to \chi$ and $\psi \to -\psi$, $y$-parity symmetry: $x \to x $, $y \to -y$ with $\chi \to \chi$, $\psi \to -\psi$ as well as charge conjugation (or particle-hole) symmetry: $\chi \to \chi^\dagger$, $\psi \to \psi^\dagger$.  Note that in this model $x$ and $y$ coordinates are treated differently as we see in $H_{\mathrm{int}}$, and there is no  $\pi/2$ rotation symmetry (which is sometimes common in a fracton phase of matter \cite{Seiberg:2020bhn}). Our supercharge and hence Hamiltonian is stable in the sense that there is no other term consistent with the above symmetry without adding more derivatives.

Our model may be regarded as a two-dimensional generalization of the extended Nicolai model \cite{Sannomiya:2016wlz} equipped with the fermionic subsystem symmetry. In the $g=0$ limit, it may be regarded as a two-dimensional generalization of the original Nicolai model.

We would like to note here that $g^{-1}$ plays the role of the coupling constant of the model. To see this, we rescale the time $t \to \frac{t}{g}$ and observe that the action is equipped with the canonical kinetic terms with the four-fermi interaction proportional to $g^{-1}$:
\begin{align}
I = \int d^2x\, dt \left( i\chi^\dagger \partial_t \chi + i \psi^\dagger \partial_t \psi  {-} (\chi^\dagger\partial_x \partial_y \psi^\dagger - \chi \partial_x \partial_y \psi)  - g - g^{-1} H_{\mathrm{int}} \right) \ . 
\end{align}
In the infinite $g$ limit (i.e., free limit), the dispersion relation of $\psi$ and $\chi$ is $E= |k_x k_y|$ in terms of the wave nunmber $k_x$ and $k_y$, and we have a ground-state degeneracy with the ``Area-law" entropy (i.e., $S \propto \mathrm{Area}$).\footnote{By convention, ``Area law" in $d$ space dimensions means that the entropy is proportional to $L^{d-1}$, where $L$ is the system size. In our $d=2$ dimensional case, ``Area law" actually means peripheral law $S  \propto L_x + L_y$ (or in our case $S \propto L_x$) as we will see in the next section.} Around infinite $g$, naive dimension counting suggests that $g^{-1}$ is irrelevant in the renormalization group sense.

Supersymmetry ($Q$ and $Q^\dagger$) is spontaneously broken when $g\neq 0$. 
For large enough $g$, the Hamiltonian is dominated by the constant term $\int d^2x\, g^2$ and the ground-state energy cannot be zero. From the anti-commutation relation $\{Q^\dagger, Q\} = H$, if the vacuum energy is non-zero, the supersymmetry must be spontaneously broken \cite{Witten:1981nf}. By using the holomorphy argument, we expect that the phase transition only occurs at $g=0$  if any. Indeed, we can formally argue for the spontaneously breaking of the supercharge $Q$ in the following way. 

Let us introduce a local operator $O^\dagger$:
\begin{align}
O^\dagger = \chi^\dagger \left( 1 -\frac{1}{g} (\partial_x \chi \partial_y \psi + \partial_x (\chi \partial_y\psi)) + \frac{4}{g^2} \chi \partial_x \chi \partial_y \psi \partial_x \partial_y \psi) \right)  \label{local}
\end{align}
such that $\{Q, O^\dagger \}  = g$ . This implies that the supercharge $Q$ is spontaneously broken for non-zero $g$ because otherwise $\langle 0|\{Q,O^\dagger\}|0\rangle = 0$. The argument is valid as long as the regularization and the renormalization do not spoil the supersymmetry transformation \eqref{local}. We will discuss a lattice regularization in the next section.

The zero-energy excitation (from the dispersion relation $E = |k_x k_y|$ in the infinite $g$ limit) can be understood as a Nambu-Goldstone fermion $\chi$ (or more precisely $O$ introduced above) with $k_x = k_y = 0$. On the other hand, the zero-energy mode of $\psi$ with $k_y= 0, k_x \neq 0$  corresponds to Goldstone fermions for the fermionic subsystem symmetry \eqref{subsystem}. The two distinct Nambu-Goldstone fermions coincide at $k_x=k_y =0$ due to the kinetic mixing  between $\chi$ and $\psi$. The zero-energy mode of $\psi$ with $k_x=0 , k_y \neq 0$ is a consequence of the emergent fermionic subsystem symmetry \eqref{dipole} in the free limit and does not survive once we introduce the interaction with finite $g$.

For finite $g$, the interaction will gap some of the zero-energy excitations that are not protected by the fermionic subsystem symmetry. The zero-energy mode protected by the fermionic subsystem symmetry is the $k_y=0$ mode. The ground-state entropy is smaller but it is still proportional to $L_x$ but is insensitive to $L_y$ (once we put the theory on a finite box with length $L_x$ and $L_y$).

When $g=0$, the model may be regarded as a variant of the Nicolai model \cite{Nicolai:1976xp} and we conjecture that the ground states are much more degenerate: the ground-state entropy shows the ``Volume law": $S \propto \mathrm{Volume} \sim L_x L_y$. One supersymmetric ground state is the Fock vacuum: $\psi(\vec{x})|\Omega\rangle = \chi(\vec{x}) |\Omega \rangle =0$, but there are many others.

It is trivial to generalize our model to three dimensions $(t,\vec{x}) = (t,x,y,z)$:
\begin{align}
Q  &= \int d^3x\left( g \chi + \chi  \partial_x \chi \partial_y \partial_z \psi \right) \ ,\cr
H &= Q^\dagger Q + Q Q^\dagger \cr
& = \int  d^3x \left( g^2  {+} g\left( \chi^\dagger \partial_x \partial_y\partial_z \psi^\dagger - \chi \partial_x \partial_y\partial_z \psi \right) + H_{\mathrm{int}} \right) 
\end{align}
with
\begin{align}
H_{\mathrm{int}} = -\int d^3x\,& \Big( \partial_y\partial_z (\chi^\dagger \partial_x \chi^\dagger) \partial_y\partial_z (\chi \partial_x \chi) + (\partial_x \chi \partial_y\partial_z \psi )(\partial_x \chi^\dagger \partial_y\partial_z \psi^\dagger) \cr 
&+ \partial_x(\chi \partial_y\partial_z \psi )(\partial_x \chi^\dagger \partial_y\partial_z \psi^\dagger) + \partial_x (\chi^\dagger \partial_y\partial_z \psi^\dagger)(\partial_x \chi \partial_y\partial_z \psi) \cr 
&+ \partial_x(\chi \partial_y\partial_z \psi) \partial_x (\chi^\dagger \partial_y\partial_z \psi^\dagger)
\Big) \ .
\end{align}
As in two dimensions, the supersymmetry $Q$ is spontaneously broken when $g\neq 0$.

The corresponding action is
\begin{align}
I = \int d^3x\, dt \left( i\chi^\dagger \partial_t \chi + i \psi^\dagger \partial_t \psi -g^2 {-} g\left( \chi^\dagger \partial_x \partial_y\partial_z \psi^\dagger - \chi \partial_x \partial_y\partial_z \psi \right) - H_{\mathrm{int}} \right) \ , 
\end{align}
and it has the fermionic subsystem symmetries generated by 
\begin{align}
q^\dagger_y(z,x) &= \int dy\, \psi^\dagger(\vec{x}) \ ,  \cr 
q^\dagger_z(x,y) &= \int dz\, \psi^\dagger(\vec{x}) \ .
\end{align}

Morally speaking, the free limit of the resulting theory is a fermionic part of the supersymmetric model studied in \cite{Yamaguchi:2021qrx} with the dispersion relation $E = |k_x k_y k_z|$ (except that our fermions are complex rather than real). 
We can argue that for finite $g$, the ground-state entropy shows the ``Area law": $S\sim L^2$. When $g=0$, we conjecture that the ground-state entropy shows the ``Volume law": $ S \sim L^3$.

\section{Lattice model and ground-state degeneracy}

One lattice realization of our continuum model is to consider the following supercharge on the square lattice:
\begin{align}
Q = \sum_{i \in \mathbb{Z}^2} \left( g c_i - c_i c_{i+\hat{x}}(d_i - d_{i+\hat{y}}) \right)  \ .   
\end{align}
Here, $i$ labels the sites and $c_i$ and $d_i$ are complex fermion operators with canonical anti-commutation relations: $\{c_i, c^\dagger_j\} = \{d_i,d^\dagger_j\} = \delta_{ij}$ (and $\{c_i,c_j\} = \{d_i,d_j\} = \{c_i,d_j \} = \{c_i,d_j^\dagger \} = 0$). In the continuum limit, $c_i$ and $d_i$, respectively, correspond to $\chi(\vec{x})$ and $\psi(\vec{x})$ in the previous section. Here and hereafter, $i + n\hat{x} + m\hat{y}$ denotes the lattice site that is translated by $n$ lattice units in $x$ direction and $m$ lattice units in $y$ direction from site $i$.

The Hamiltonian is given by the anti-commutator:
\begin{align}
H =& Q^\dagger  Q + Q Q^\dagger \cr
 =& \sum_{i \in \mathbb{Z}^2} \left( g^2  - g(c^\dagger_{i} - c^\dagger_{i+\hat{x}})(d_i^\dagger - d^\dagger_{i+\hat{y}}) +  g(c_{i}-c_{i+\hat{x}})(d_i-d_{i+\hat{y}})+ H_{\mathrm{int}} \right)  \ , 
\end{align}
where the interaction Hamiltonian $H_{\mathrm{int}}$ is 
\begin{align}
H_{\mathrm{int}} 
=&\sum_{i \in \mathbb{Z}^2} \Big( - c^\dagger_{i+\hat{x}} c_{i+\hat{x}} (d_i -d_{i+\hat{y}})(d^\dagger_i -d^\dagger_{i+\hat{y}}) + c^\dagger_{i} c_{i+2\hat{x}}(d_{i+\hat{x}} - d_{i+\hat{x}+\hat{y}})(d^\dagger_{i}-d^\dagger_{i+\hat{y}}) \cr
&-c_i c^\dagger_{i+2\hat{x}}(d_{i} - d_{i+\hat{y}})(d^\dagger_{i+\hat{x}}-d^\dagger_{i+\hat{x}+\hat{y}}) + c_i c^\dagger_i (d_i -d_{i+\hat{y}})(d^\dagger_i -d^\dagger_{i+\hat{y}}) \cr
& + c^\dagger_{i+\hat{x}} c^\dagger_i c_i c_{i+\hat{x}}- c^\dagger_{i+\hat{x}+\hat{y}}c^\dagger_{i+\hat{y}} c_{i}c_{i+\hat{x}} -c^\dagger_{i+\hat{x}-\hat{y}}c^\dagger_{i-\hat{y}} c_{i}c_{i+\hat{x}} + c^\dagger_{i+\hat{x}} c^\dagger_i c_i c_{i+\hat{x}} \Big) \ . 
\end{align}
A similar supercharge calculation can be found in \cite{Sannomiya:2016wlz}, where the extended Nicolai model was studied.

As in the continuum field theory, we can argue for the spontaneous symmetry breaking of $Q$ by the trick used in \cite{Sannomiya:2016wlz,Moriya:2018fgr}. Let us introduce the local operator $O_i^\dagger$:
\begin{align}
O^\dagger_i = c_i^\dagger& \Big( 1 + \frac{1}{g}\, c_{i+\hat{x}}(d_i - d_{i+\hat{y}}) - \frac{1}{g}\, c_{i-\hat{x}}(d_{i-\hat{x}}-d_{i-\hat{x}+\hat{y}}) \cr
&+ \frac{2}{g^2}\, c_{i+\hat{x}}c_{i-\hat{x}}(d_i-d_{i+\hat{y}})(d_{i-\hat{x}}-d_{i-\hat{x}+\hat{y}}) \Big)
\end{align}
such that $\{Q,O_i^\dagger \} = g$. This immediately implies that the supersymmetry $Q$ is spontaneously broken unless $g=0$ on the finite lattice. A careful analysis of the infinite volume limit may be completed as in \cite{Moriya:2018fgr}.

In the free limit $g \to \infty$, the dispersion relation becomes $E = 4|\sin \frac{ k_x}{2} \sin\frac{k_y}{2}|$ (after the rescaling of time $t \to g^{-1}t$). The gapless mode at $k_x=k_y =0$ corresponds to a Nambu-Goldstone fermion for the spontaneously broken supersymmetry $Q$ (when $g \neq0$). 

Let us now study the fermionic subsystem charges and ground-state degeneracy on the finite lattice. We consider a square lattice with the lattice size $L_x \times L_y$  (i.e. $i = i + L_x \hat{x} = i + L_y\hat{y} $), imposing periodic boundary conditions on fermions.

We can clearly see that the fermionic subsystem charges
\begin{align}
q_m &= \frac{1}{\sqrt{L_y}} \sum_{n \in \mathbb{Z}_{L_y}} d_{m \hat{x} + n\hat{y}} 
\end{align}
commute with the Hamiltonian, and they satisfy the standard anti-commutation relations of $L_x$ complex fermions:
\begin{align}
\{q_m, q_{m'}^\dagger \} &= \delta_{m,m'} , \ \  \{q_m,q_{m'} \} = 0 \ .
\end{align}
The ground states form a non-trivial representation of this algebra so that the ground-state entropy shows $S \sim {L_x}$ at least. When $g=\infty$ it is enhanced to $S \sim {L_y+L_x}$ due to the extra zero modes from $\sum_{m} d_{m{\hat{x}}+n\hat{y}}$. When $g=0$, our model becomes a variant of the Nicolai model and we conjecture that the ground-state entropy is given by $S\sim {L_x L_y}$. 

We can estimate a lower bound of the ground-state entropy at $g=0$ as follows. First note that if we take any of the (classical) supersymmetric ground states of the Nicolai chain (see e.g. \cite{Katsura:2017pjd}) copied in $y$ directions are the supersymmetric ground states of our two-dimensional model (leading to $S \sim L_x$ ground-state entropy). Now we may still obtain the supersymmetric ground states by flipping some of the occupancies of $c_i$ while keeping the $d_j$ states. If this happens to be the case, we can repeat the flipping arbitrarily many times along the $c_i$ in the $y$ directions. This procedure is local in the $x$ direction so we obtain $S \sim {L_x L_y}$ entropy.

We offer numerical studies of the ground-state degeneracy with small lattices. With $(L_x,L_y) = (2,2)$, we have $160$ ground states at $g=0$. With $(L_x,L_y) = (3,2)$, we have $1504$ ground states at $g=0$. Similarly, with $(L_x,L_y) = (2,3)$, we have 1792  ground states at $g=0$.


When $g\neq0$, we may give a more precise conjecture about the ground-state degeneracy. For $L_x >2$, it is conjectured to be $4 \cdot 2^{L_x} $. The number can be explained as follows. We have $L_x$ zero modes from $\sum_{n} d_{m\hat{x}+n\hat{y}}$ as well as one zero mode from the spontaneously broken supercharge $Q$, leading to the $2^{L_x +1}$-fold degeneracy. We further observe that the charge conjugation $Q \to Q^\dagger$ acts non-trivially on them so that we have the $2^{L_x +2}$-fold degeneracy in total.\footnote{The extra degeneracy from the charge conjugation had already appeared in the one-dimensional extended Nicolai chain \cite{talk}. Suppose that the charge conjugation acts on a ground state $|\Omega \rangle$ trivially (otherwise the extra degeneracy is obvious), then we can always find a quartet $|\Omega \rangle, Q|\Omega \rangle, Q^\dagger |\Omega \rangle$ and $ Q Q^\dagger |\Omega \rangle (=(E-Q^\dagger Q)|\Omega \rangle)$ as long as the supersymmetry is spontaneously broken. One can then rearrange the quartet into a pair $Q^\dagger |\Omega \rangle$ and $QQ^\dagger|\Omega \rangle$ and their charge conjugates. The argument is generic and it applies to any spontaneously broken supersymmetric models with the charge conjugation symmetry (in finite systems).}  

For $L_x=2$, the situation is slightly different. We have $L_y$ zero modes from $\sum_{m} d_{m\hat{x}+n{\hat{y}}}$ and two zero modes from $\sum_{n} d_{m\hat{x}+n\hat{y}}$, but there is one trivial relation $\sum_n (\sum_m  d_{m\hat{x}+n{\hat{y}}}) = \sum_m (\sum_n  d_{m\hat{x}+n{\hat{y}}})$ among them. In addition, the supercharge $Q$ gives another zero mode, leading to the $2^{L_y +2}$-fold degeneracy. We observe that the charge conjugation $Q \to Q^\dagger$ acts non-trivially (at least  when $L_y = 4, 5$) so that we have the $2^{L_y+3}$-fold degeneracy in total. 
 
 We claim that there are no other zero modes that are not associated with any symmetry. We have less evidence for $L_x \neq 2$ and $L_y \neq 2$, but for small numbers of $L_x$ and $L_y$ we have explicitly checked with a computer that there are no other zero modes. The above counting does not depend on $g$ (as long as $g\neq 0$), and we do not expect any phase transition at finite $g$.

Similarly, in three dimensions, we consider the supercharge on the cubic lattice:
\begin{align}
Q = \sum_{i \in \mathbb{Z}^3} \left( g c_i  + c_{i} c_{i+\hat{x}}(d_{i} - d_{i+\hat{y}} - d_{i+\hat{z}} + d_{i+\hat{y}+\hat{z}}) \right) \ ,
\end{align}
and define the Hamiltonian by the anti-commutator $H = \{Q^\dagger, Q\}$. We can study the degeneracy of the ground states explicitly and verify the ``Area-law" entropy when $g \neq 0$. For example, when $L_x = L_y = L_z =2$ the degeneracy is $256$ both at $g=0$ and $g=1$.\footnote{We have seven $\psi$ zero modes as well as one zero mode from the supercharge, resulting in the $2^{8}$-fold degeneracy. This does not depend on $g$. On larger lattices, we expect that the degeneracy depends on whether $g=0$ or not.} When $g=0$, we conjecture that the ground-state degeneracy gives the ``Volume law" on a larger lattice.

\section{Other models}
Here, we discuss other models that have similar features to our models studied above.

The first model is to construct a supercharge out of one fermion $\chi$ per one lattice unit (instead of two fermions $\chi$ and $\psi$ per one lattice unit).
The supercharge is given by 
\begin{align}
Q = \int d^2x \left( g \chi + \chi \partial_x \chi \partial_y \chi \right) \ 
\end{align}
with the Hamiltonian $H = \{Q^\dagger, Q\}$. Note that this model does not possess the $U(1)$ symmetry when $g\neq 0$.
In the infinite $g$ limit, the dispersion relation is $E =|k_x k_y|$ and the ground-state degeneracy shows the ``Area-law" entropy. On the other hand, with finite but non-zero $g$, while the supersymmetry $Q$ is still spontaneously broken, the ground-state degeneracy is lifted because the interaction lacks the subsystem symmetry. One lattice realization of this model on the triangular lattice was studied in Appendix F of \cite{Sannomiya:2016mnj}.

The second model is to restore the asymmetry between $x$ and $y$ directions. For this purpose, we may introduce two $\psi$ fields $\psi_1$ and $\psi_2$. Then the supercharge
\begin{align}
Q = \int d^2x \left( g \chi + \chi \partial_x \psi_1 \partial_y \psi_2 \right)
\end{align}
with the Hamiltonian $H = \{Q^\dagger, Q \}$ gives the model that retains the $\pi/2$ rotation symmetry (under which $\psi_1$ and $\psi_2$ are non-trivially exchanged). In the infinite $g$ limit, the dispersion relation is $E = |k_x k_y|$ (from $\psi_1$ and $\psi_2$ modes) together with a flat band at zero energy (from $\chi$ modes). With finite $g$, the flat band is lifted and we conjecture that the ground-state degeneracy shows the ``Area-law" entropy because of the fermionic subsystem symmetry $\psi_1(\vec{x}) \to \psi_1(\vec{x}) + \zeta(y)$ and $\psi_2(\vec{x}) \to \psi_2(\vec{x}) +  \xi(x)$.

The third model is to achieve the full fermionic subsystem symmetry in both $x$ 
and $y$ directions: $\psi(\vec{x}) \to \psi(\vec{x}) + \xi(x) + \zeta(y)$. For this purpose, we introduce two $\chi$ fields $\chi_1 $ and $\chi_2$ and consider the supercharge
\begin{align}
Q = \int d^2x  \left(g \chi_1 + \chi_1 \chi_2 \partial_x \partial_y \psi \right) \  
\end{align}
with the Hamiltonian $H = \{Q^\dagger, Q \}$. It has the fermionic subsystem symmetry  $\psi(\vec{x}) \to \psi(\vec{x}) + \xi(x) + \zeta(y)$. In the infinite $g$ limit, it shows the dispersion relation $E = |k_x k_y|$ (from $\chi_2$ and $\psi$ modes) with an additional flat band at zero energy (from $\chi_1$ modes). With finite but non-zero $g$, we conjecture that the ground-state degeneracy shows the ``Area-law" entropy.

\section{(Anomalous) Gauging and defect operators}

Our models discussed so far do not possess a fracton as a dynamical degree of freedom, but we may consider the analogue of a probe fracton by considering a defect in the gauged theory. Our gauge theory will be a fermionic analogue of the tensor gauge theory studied in the context of a fracton phase of matter \cite{You:2019cvs,You:2019bvu} (see also \cite{Tantivasadakarn:2020lhq,Shirley}). 

We would like to gauge the shift symmetry of $\psi$ fields by replacing ordinary derivatives $\partial_t$ and $\partial_y$ with covariant derivatives in the action. In the spatial direction, it is not difficult. We simply  replace $\partial_y \psi$ with the covariant derivative $D_y \psi = \partial_y \psi + \Phi_y$ by assuming the fermionic gauge transformation law:
\begin{align}
\psi(t,\vec{x}) &\to \psi(t,\vec{x}) + \lambda(t,\vec{x}) \ , \cr
\chi(t,\vec{x}) & \to \chi(t,\vec{x}) \ , \cr
\Phi_t(t,\vec{x}) & \to \Phi_t(t,\vec{x}) - \partial_t \lambda(t,\vec{x}) \ , \cr 
\Phi_y(t,\vec{x}) & \to \Phi_y(t,\vec{x}) - \partial_y \lambda(t,\vec{x}) \ .
\end{align}
Here the gauge fields $\Phi_t$ and $\Phi_y$ as well as the gauge parameter $\lambda$ are fermionic.

On the other hand, gauging of the shift symmetry with the first order time derivative requires extra care. We will see that gauging of our model itself is anomalous in the t'Hooft sense and cancellation of the anomaly is necessary to make the gauge field dynamical. 
The difficulty comes from the fact that the action is invariant under the shift only up to total derivatives in time.\footnote{In this section, we will address this issue in the field theory language, but it has the same difficulty with the lattice regularization. In Appendix B we will discuss the origin of the problem in the simplest quantum mechanics in zero spatial dimensions.}
 
Let us postulate the action of the form
\begin{align}
\int d^2x\, dt\, (i\psi^\dagger D_t \psi) = \int d^2x\, dt\, ( i\psi^\dagger \partial_t \psi + i\Phi^\dagger_t \psi - i\psi^\dagger\Phi_t) \ 
\end{align}
to describe a gauged kinetic term. The variation of the gauge fields $\Phi_t$ cancel the first order variation of $\psi$ fields, but we see that the action fails to be invariant because of the remainder: 
\begin{align}
\int d^2x\, dt\, i(\Phi^\dagger_t  \lambda - \lambda^\dagger \Phi_t)  \label
{anomaly}
\end{align}
which cannot vanish for non-zero $\Phi_t$, which we call ``anomaly" (because it only depends on the gauge fields rather than the matter field $\psi$). The counter-term that could be used to cancel the anomaly would be $i\Phi_t^\dagger \partial_t^{-1} \Phi_t$, but it is non-local. 

Therefore, in order to gauge the fermionic shift symmetry, we need a certain cancellation mechanism by adding other degrees of freedom. In this discussion, we did not assume the supersymmetry, but we will see the same difficulty in the full supersymmetric action in Appendix A. The only difference is that the gauge field must have a reality condition $\lambda^\dagger=-\lambda$.

Under the assumption of the anomaly cancellation, we may study the defect line operator that would correspond to a fermionic ``fracton" or ``lineon"\cite{Pretko:2020cko}.
We can construct a gauge invariant fermionic defect line operator
\begin{align}
\Psi = \int \left( dt\, \Phi_t + dy\, \Phi_y \right) \ .
\end{align}
The integration is over the worldline trajectory of the probe.
Since we do not have the gauge field $\Phi_x$, the defect line operator cannot ``move" in  $x$ direction (i.e., the probe trajectory cannot possess the non-trivial $dx$ component). This defect operator becomes an analogue of ``fracton" or ``lineon" in our model. 

\section{Discussions}
In this paper, we have constructed a purely fermionic system with spontaneously broken supersymmetry that shares the common feature with a fracton phase of matter. Our model is gapless due to the Nambu-Goldstone mechanism. It shows a ground-state degeneracy with the ``Area-law" entropy due to the fermionic subsystem symmetries.  In the strongly coupled limit, it becomes a variant of the Nicolai model, and the ground-state degeneracy shows the ``Volume-law" entropy.

Our model is characterized by a coupling constant $g$. The phase boundary of the spontaneous supersymmetry breaking is located at $g=0$. This is expected from the holomorphic nature of the supersymmetric model with a complex supercharge. It may be interesting to consider the Majorana version of our model so that the phase transition at finite $g$ is possible. On the one dimensional chain, we find that there exists a phase transition at finite $g$ \cite{Sannomiya:2017foz}.

From a formal perspective, one might have noticed that the subsystem symmetry in our model is rather trivially realized as a linear combination of fundamental fermionic operators $d_i$  (or field $\psi$). One may even try to remove these zero-mode states from the model and the exponential degeneracy of the ground states would be gone. We note that the apparent simplicity of removing the zero mode associated with the subsystem symmetry is almost always the case in the effective field theory description \cite{PBF,Seiberg:2020bhn,Seiberg:2020wsg,Yamaguchi:2021qrx,Burnell:2021reh,Gorantla:2021bda}. In the continuum description, however, removing the zero modes leads to a non-locality and break down of the effective field theory. 

For future directions, it seems important to understand the anomaly associated with gauging of the fermiomic shift symmetry. We have demonstrated it from the continuum field theory perspective, but understanding of it from the lattice viewpoint may be of interest. It may be related to the difficulty to construct a model of fermionic fracton phase of matter.

The study of the fracton effective field theories is still in its infancy. Non-trivial UV/IR mixture may make the study of the interaction and the renormalization group non-trivial \cite{Gorantla:2021bda,Distler:2021bop}. Realization in string theory and holography may help better understand the strongly interacting phase \cite{Geng:2021cmq} and reveal a possible connection to the black hole physics through the ``Area-law" entropy formula.

\section*{Acknowledgements}

This work by YN is in part supported by JSPS KAKENHI Grant No. 21K03581. 
This work by HK is in part supported by JSPS Grant-in-Aid for Scientific Research on Innovative Areas No. JP20H04630, JSPS KAKENHI Grant No. JP18K03445, Grant-in-
Aid for Transformative Research Areas (A) ``Extreme Universe" No. JP21H05191[D02], and the Inamori
Foundation.

\appendix
\section{Superfield formalism}
To obtain the supersymmetric action, we may use the superfield technique (see e.g. \cite{Witten:1981nf}).\footnote{Our superspace formalism is much simpler than the one used in \cite{Yamaguchi:2021qrx}, where the anti-commutator of the supercharge also involves the spatial derivatives.} Let us represent the supersymmetry by superspace translation:
\begin{align}
Q &= \frac{\partial}{\partial \theta} +i \bar{\theta}\frac{\partial}{\partial t} \ , \cr
\bar{Q} &= \frac{\partial}{\partial \bar{\theta}} +i \theta \frac{\partial}{\partial t} \ . 
\end{align}
Let us introduce superderivative 
\begin{align}
D &= \frac{\partial}{\partial \theta} -i \bar{\theta}\frac{\partial}{\partial t} \ , \cr
\bar{D} &= \frac{\partial}{\partial \bar{\theta}} -i \theta \frac{\partial}{\partial t} \ , 
\end{align}
with fermionic chiral superfields $\bar{D} \Psi$ = 0 and $\bar{D} X = 0$. 
In component, they are expressed by
\begin{align}
\Psi &= \psi + \theta f_{\psi} - i\theta \bar{\theta} \partial_t\psi \ , \cr
X  &= \chi + \theta f_{\chi} - i\theta \bar{\theta} \partial_t \chi \ ,
\end{align}
where $f_{\chi}$ and $f_{\psi}$ are bosonic (auxiliary) fields.

The supersymmetric action can be constructed as superspace integration by noting $\theta \bar{\theta}$ component of a general superfield and $\theta$ component of a chiral superfield is supersymmetric (up to total derivatives in time). In our model, it is given by
\begin{align}
I =&  \frac{1}{2}\int d^2x\, dt\, d\theta\, d\bar{\theta}\, \left(\bar{\Psi}\Psi + \bar{X} X \right) + \frac{1}{\sqrt{2}}\left( \int d^2x\, dt\, d\theta\, (g X + X \partial_x X \partial_y \Psi) + h.c. \right) \cr
=& \int d^2x\, dt\, \Big(i \bar{\psi} \partial_t \psi + i \bar{\chi}\partial_t \chi + \frac{1}{2}(\bar{f}_{\psi}f_{\psi} + \bar{f}_{\chi} f_{\chi})  \cr
&+ \frac{1}{\sqrt{2}}( g f_{\chi} + f_{\chi} \partial_x \chi \partial_y \psi -\chi \partial_x f_{\chi} \partial_y\psi + \chi \partial_x \chi \partial_y f_{\psi} + h.c. ) \Big)  \ .
\end{align}
We can integrate out $f_{\psi}$ and $f_\chi$ to obtain the Hamiltonian
\begin{align}
H =\left( g + \partial_x \chi \partial_y \psi + \partial_x(\chi \partial_y\psi) \right) \left( g - \partial_x \chi ^\dagger \partial_y \psi^\dagger - \partial_x(\chi ^\dagger\partial_y\psi^\dagger) \right)  - \left( \partial_y (\partial_x \chi \chi^\dagger))(\partial_y (\partial_x \chi^\dagger \chi) \right) \ .  \label{last}
\end{align}

In order to gauge the shift symmetry in a manifestly supersymmetric fashion, we introduce a fermionic real superfield $V$ and a fermionic chiral superfield $\Xi_y$ and impose invariance under the superfield gauge transformation
\begin{align}
\Psi &\to \Psi + \Omega \ , \cr
X  &\to X \ , \cr
V &\to V - \Omega - \bar{\Omega} \ , \cr
\Xi_y &\to \Xi_y - \partial_y \Omega 
\end{align}
with respect to a fermionic chiral superfield $\Omega$, which plays a role of the gauge parameter. 
Then the supersymmetric matter action
\begin{align}\int d\theta\, X \partial_x X (\partial_y\Psi + \Xi_y)  + h.c.
\end{align}
is superfield gauge invariant.

We would like to discuss the supersymmetric analogue of the t'Hooft anomaly mentioned in section 5. Consider the candidate supersymmetric action
\begin{align}
\int d^2x\, dt\, d\theta\, d\bar{\theta}\, \frac{1}{2} ( \bar{\Psi}\Psi + V\Psi + \bar{\Psi}V ) \ , \label{superanomaly}
\end{align}
which would give a kinetic term.
This is superfield gauge invariant up to the anomaly term $V\Omega + \bar{\Omega} V$, but without extra degrees of freedom, there is nothing we can add to cancel it. Note in particular that $V^2 = 0$ since $V$ is fermionic.\footnote{This should be contrasted with a theory with a bosonic superfield where $\int d\theta\, d\bar{\theta}\, (\Phi+\bar{\Phi}+V)^2$ is a perfectly good action which gauges the shift symmetry of $\Phi$.}

In order to partially fix the gauge symmetry induced by $\Omega$, while assuming the anomaly cancellation, we may use the Wess-Zumino gauge with the component
\begin{align}
V &= -i\Phi_t \theta\bar{\theta} \ , \cr
\Xi_y & = \frac{1}{2}\Phi_y + a_y \theta - i\theta \bar{\theta} \partial_t \frac{1}{2}\Phi_y \ .
\end{align}
This fixes the real part of the top component of $\Omega$ (i.e. $\theta$ independent component of $\Omega$) and $\theta$ component of $\Omega$.
We still have the residual gauge transformation
\begin{align}
\psi &\to \psi + \lambda \ , \cr
\Phi_t &\to \Phi_t - \partial_t \lambda \ , \cr
\Phi_y &\to \Phi_y - \partial_y \lambda \ .
\end{align}
Here we note that $\Phi_t$ is pure imaginary while $\Phi_y$ is complex. The fermionic gauge parameter $\lambda$ is pure imaginary. In component, \eqref{superanomaly} gives the supersymmetric version of the anomalous action that we studied in section 5.

In the superspace language, the supersymmetric defect line operator is given by
\begin{align}
\Psi = \int dt\, d\theta\, d\bar{\theta}\, V  + \int dt\, d\theta\, d\bar{\theta}\, dy\, \Xi_y \ . \end{align}
Note that our superspace does not involve spatial coordinates, so it preserves the same amount of supersymmetry as the bulk action.




\section{Gauging bosonic/fermionic shift symmetry in quantum mechanics}
In this appendix, we will demonstrate the anomaly of gauging a fermionic shift symmetry in the simple quantum mechanics in comparison with a bosonic shift symmetry. Let us begin with a bosonic case. Consider a trivial particle quantum mechanics (in zero spatial dimension) with the action 
\begin{align}
I = \int dt\,  (p \partial_t x) 
\end{align}
that gives the zero Hamiltonian. The equal time commutation relation is $[x,p]= i$. It has a shift symmetry of $p \to p + \mathrm{const}$ and $x \to x+\mathrm{const}$.

If we tried to gauge the shift of {\it both} $x$ and $p$, we would obtain the gauged action
\begin{align}
I= \int dt\, (p \partial_t x + x A_p + p A_x) \ , 
\end{align}
which is anomalous. This is because the shift symmetry is a symmetry of the Lagrangian only up to total derivatives in time. To see this in the Hamiltonian formulation, we observe that 
the Hamiltonian becomes non-zero $H= -xA_p -p A_x$ while the gauge constraints are $x=0$ and $p=0$. They are, however, inconsistent with the Hamiltonian dynamics because $i[x,H] = A_x$ and $i[p,H] = -A_p$ unless $A_x = A_p =0$. This is a manifestation of the anomaly in the Hamiltonian framework. 

Instead, if we only gauge the shift of $x$, the action is {\it not} anomalous:
\begin{align}
I= \int dt (p \partial_t x + p A_x) \ . 
\end{align}
The Hamiltonian is $H=-p A_x$ and in this case we see that the gauge constraint $p=0$ is consistent with the Hamiltonian dynamics. In this way, we may be able to gauge the shift symmetry of the bosonic system, but only one of the shift of $x$ or $p$ can be gauged.

The fermionic system has a drastic difference. Consider a complex fermion with the action
\begin{align}
I = \int dt\, (i\psi^\dagger \partial_t \psi) \ , 
\end{align}
which implies the zero Hamiltonian. The equal-time anti-commutation relation (in zero spatial dimension) is $\{\psi^\dagger, \psi\} = 1$. It has a shift symmetry of $\psi \to \psi + \mathrm{const}$.
We will see that gauging the shift symmetry of $\psi$ is anomalous. 

Without loss of generality, let us try to gauge the shift of $\psi + \psi^\dagger$. Would-be gauged action becomes 
\begin{align}
I= \int dt (i \psi^\dagger \partial_t \psi + i\Psi (\psi + \psi^\dagger) \ , 
\end{align}
but the gauge transformation is anomalous. The Hamiltonian is $H= -i\Psi (\psi + \psi^\dagger)$, but the gauge constraint is inconsistent with the Hamiltonian dynamics $i[H, \psi+\psi^\dagger] = 2\Psi$ unless $\Psi = 0$. 

Our discussion is generic, but there is a possible loophole: one may add extra degree of freedom and/or modify the gauge transformation of $\Psi$. One successful example of gauging a fermionic shift symmetry is the super-Higgs mechanism in supergravity, where the space-like component of the gravitino and the gravity play the role of the extra degrees of freedom and the modification of the gauge transformation is accompanied \cite{Deser:1977uq}.

\end{document}